\documentclass[a4paper,english]{article}
\usepackage[T1]{fontenc}
\usepackage[utf8]{luainputenc}
\usepackage{graphicx}

\makeatletter

\pdfpageheight\paperheight
\pdfpagewidth\paperwidth

\newcommand{\lyxaddress}[1]{
\par {\raggedright #1
\vspace{1.4em}
\noindent\par}
}

\makeatother

\usepackage{babel}
\begin{document}

\title{\noindent A Bi-Directional Big Bang\,/\,Crunch Universe \\ within
a Two-State-Vector Quantum Mechanics?}

\author{Fritz W. Bopp}
\maketitle
\vspace{-6.6em}
\lyxaddress{\noindent \begin{center}
Department of Physics, Siegen University, Germany
\par\end{center}}
\begin{abstract}
\noindent {\normalsize{}A two boundary quantum mechanics incorporating
a big bang / big crunch universe is carefully considered. After a
short motivation of the concept we address the central question how
a proposed a-causal quantum universe can be consistent with what is
known about macroscopia and how it might find experimental support.}{\normalsize \par}
\end{abstract}

\lyxaddress{Keywords: Two state vector interpretation of quantum mechanics, resurrection
of macroscopic causality, big bang / big crunch universe}

~\\
In the literature searching for consistent interpretation of quantum
mechanics the Einstein, Podolsky and Rosen ,,paradox``~\cite{einstein1935can}
plays a dominant role. Many solutions were proposed. There are two
main options. One can change or limit the ontology of wave functions
(or fields) like it is done in the old Copenhagen interpretation (see
p.e.~\cite{bohr1935can,bopp2013werner}) or one can admit retro-causation
(see p.e.~\cite{Argaman2010Bell,wheeler1949classical,de1953mechanique,cramer1986transactional,price2012does}). 

In our view there is actually a more decisive ,,paradox`` of quantum
statistical nature. We are at $\tau_{0}=0$. Consider two wavelets
(or suitably regulated point fields) of identical particles created
around the time $\tau_{1}<\tau_{0}$ and $\tau_{2}<\tau_{0}$ in areas
within our backward light cone and annihilated around the times $\tau_{3}>\tau_{0}$
and $\tau_{4}>\tau_{0}$ in areas within our forward cone. Considering
the essential part $a(\tau_{1})\, a(\tau_{2})\, a^{+}(\tau_{3})\, a^{+}(\tau_{4})$
one obtains two contributions:
\[
\left[a(\tau_{1})\, a^{+}(\tau_{3})\right]\,\left[a(\tau_{2})\, a^{+}(\tau_{4})\right]\,\,\pm\,\,\left[a(\tau_{1})\, a^{+}(\tau_{4})\right]\,\left[a(\tau_{2})\, a^{+}(\tau_{3})\right]
\]
There are restriction on the functional behavior of each commutator
but the effect considered is independent of the details of their behavior. 

The probability of the creation and annihilation process depends on
the square of the amplitude and the relative phase of both contributions
enters. The point is now that this phase also depends on the Hamiltonian
for $\tau>\tau_{0}$. If the Hamiltonian is manipulated by us at $\tau_{0}$
it those impacts the $\tau<\tau_{0}$ creation probability happening
in our backward light cone. In our opinion the effect destroys the
first option (which restricts retro causation just to wave functions
not considered ontological) as here in principle observable probabilities
are involved.

In previous papers we discussed less abstract implementations of this
paradox. We carefully considered~\cite{Bopp:2016nxn} the Humbury-Brown
Twiss effect used in astronomy~\cite{brown1957interferometry} and
multi-particle physics~\cite{metzger2011bose,kittel2005soft}. We
also presented~\cite{Bopp:2018kiw} a simple gedanken experiment
with two radio antennae sitting in the focal points of a mirrored
ellipsoid possibly emitting each a single photon at $t=-\Delta T$
with a electronically chosen phase. The photons are then absorbed
at $t=+\Delta T$ at opposite sides. Everything is known and calculable.
A dark spot on the surface put just before $t=0$ allows for an extra
counting interference contribution. If in a positive region it increases
the emission probability at $t=-\Delta T$ in a manifest retro causal
way. 

In our opinion there is no escape to avoid backward causation. However
retro causation requires extremely rare conditions which can be ignored
in macroscopic considerations. 

Measurements inhabit the interface between quantum dynamics (quantum
mechanics without jumps) and macroscopia. How does retro causation
change the concept of measurements. A measurement is usually associated
with a collapse removing the other components. Formally it is represented
by a renomalized to-one-state projection operator. Retro causation
allows the measurement to happen any time after the initial splitting.
To postpone the collapse p.e. into an extended detector will not effect
the Born rule for the initial splitting. The argument is not completely
trivial as one initial option could asymmetrically lead to more candidate
states to collapse to. It follows from unitarity of the evolutions
of both components between the splitting and the measurement. 

In the coherence concept~\cite{joos2013decoherence} one considers
an open system and assumes that the measurement occurs if a witness
reaches for all practical purposes the ,,outside`` attributable
to macroscopia. We here consider a closed universe with an initial
and final state. Extending the basic idea of the coherence concept,
the measurement with its collapse is taken to occur if a witness reaches
the ,,external`` final state.

Formally the postponed measurement point can be written (ignoring
normalization) as a shift of the corresponding projection:

\[
<\mathrm{initial}\,|\, U_{1}\cdot\mathrm{Projection}\cdot U_{2}\,|=<\mathrm{initial}\,|\, U_{1}\cdot U_{2}\cdot\mathrm{Projection'}\,|
\]

Obviously this projections can then be included in the final state
density matrix:

\[
\rho\propto\sum_{i}\mathrm{Projection_{i}}'\cdot\rho_{0}\cdot(\sum_{i}\mathrm{Projection_{i}}')^{\dagger}
\]

As the number of decisions fixed by projection is huge $\rho$ has
to be very restrictive. It presumably suffices to consider a single
fixed final state:

\[
\rho=|\mathrm{final}><\mathrm{final}|
\]

and to do the same for the initial state density matrix. Both assumptions
are not crucial. 

The collapse of a measurement contain besides the projection operator
a huge re-normalization factor 

\[
\frac{1}{\left|<\mathrm{initial}\,|\,\mathrm{Projection}\,\right|}<\mathrm{initial}\,|\,\mathrm{Projection}\,
\]

and for the wave function side of the universe a measurements means:
\[
<\mathrm{initial}\,|\, U_{1}\cdot\mathrm{Projection}\cdot U_{2}\,|\,\mathrm{final}>/<\mathrm{initial}\,|\, U_{1}\,\cdot\, U_{2}\,\mathrm{|\, final}>\,.
\]

It is the two boundary formalism developed by Aharonov and coworkers~\cite{aharonov1964time,2015arXiv151206689A}
and others~\cite{gell1994time,griffiths1984consistent}. 

It is actually quite close to a multiverse interpretation~\cite{everett1957relative,Vaidman:2014noa}.
In this interpretation ,,our`` universe is determined by a community
of ,,our`` observers. Only branching is considered. At the end ,,our``
universe is one out of $2^{\mathrm{\mathrm{decisions}}}$ others.
The point is now that nothing changes for ,,our`` present situation
if shortly before its end a projection operator is entered eliminating
all other universes for the remaining time. However restricting the
consideration to ,,our`` universe it can now be described in a two
boundary way. If the projection operator factorizes it exactly corresponds
to the two boundary state situation.

It is of course not certain that the final state is reached by a decisive
witness. A trivial example is drawn in Fig.~1. A sideways polarized
electron is initially split in a up and down component. If a sufficiently
good vacuum (e.c.t.) avoids traces the shown arrangement destroys
the up/down distinction. Such coexisting intermediate paths clearly
exist in the quantum world. An obvious requirement is to eliminate
them on a macroscopic level. We assume that the witnesses reaching
the final state are structured sufficiently detailed to practically
always satisfy this requirement. 
\begin{figure}[h]
\noindent \centering{}\includegraphics[scale=0.37]{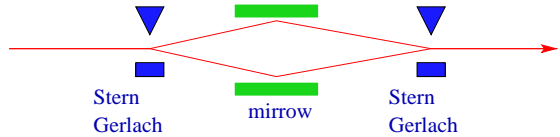}\protect\caption{The traceless Stern Gerlach splitting}
\end{figure}

Consider the Schrödinger cat. If the box containing the cat would
really be perfectly insulated and kept for $10{}^{\mathrm{huge}}$
years (or covering an entire closed grandpa-paradox temporal loop
in the framework of general relativity) the coexistence of a live
or dead cat (or grandpa) would be possible. In a finite size box all
possible states will typically be reached eventually which effectively
destroys traces in the final state and eliminates its power to prevent
coexisting macroscopic states. 

Even ignoring the time requirement the box will never be completely
insulated and thermal (or lower energy) photons will witness the macroscopic
situation. Such photons or their offspring will eventually make it
to the thin sky and so eventually reach the the end of the expanding
universe. In this way the sky plays a significant role. It replaces
the conscious observer often discussed in literature~\cite{wigner1961remarks}.
Of course the reflection-less dark sky of the expanding universe is
also the cause of the thermodynamical time arrow~\cite{zeh2001physical}.

The mechanism to fix the macroscopic situation is not simple. Tiny
interactions with a strongly self interacting system might even enhance
the stability of quantum states (discussed in~\cite{plenio2008dephasing}
for biological systems). Intense self interactions within the surrounding
system somehow limit the decisiveness of the eventual final state.
It somehow resurrects the quantum Zenon stability. The effect~\cite{balzer2002relaxationless}
is usually eliminated as the required macroscopic outcome (like an
emission of a photon in atomic physics~\cite{block1991quantum})
leads to a tide correlation of the interaction times in the evolution
of the amplitude and its conjugate eliminating one power in the dependence
on the decreasing ,,measuring`` interval essential for the effect.
For the evolution of the universe - at least after freeze out of a
plasma state - we consider such effects statistically irrelevant.

In contrast to the quantum world macroscopic considerations should
not allow for distinct coexisting path ways. A large number of effective
measurements must reduce ambiguities to allow for one macroscopic
description. In the two boundary description these measurements must
stored in the boundary states. This means that the overlap
\[
<\mathrm{inital}\,|\,\mathrm{final}>\sim0.5^{\mathrm{decisions}}
\]
estimated in~\cite{joos2013decoherence} must be tiny. 

We are aware fixing the macroscopic situation is not an easy task
for the dark sky and the huge number of needed decisions is a problematic
point in this interpretation. A single cosmic ray particle entering
the atmosphere sometimes produces thousands of particles all causing
different new macroscopic situations which somehow have to be traced. 

This motivates some authors to add a kind of dynamical jump process
to the theory. We consider it unsatisfactory as it seems at hoc~\cite{ghirardi1986unified}
or if gravitation~ is used~\cite{PenroseRoad,Gambini:2015zda} effectively
so.

Our final state has not to store the complete macroscopic situation
but just nail it down in the subspace of possibilities. With a given
macroscopic scales the number of situations in a closed universe is
not formally infinite. Also there are efficient witnesses which are
,,cheap`` and macroscopically unnoticeable. A single 30 cm radio
wave photon possibly emitted from a circuit board or the brain of
Wigner's conscious observer carries away an undetectable energy of
$10^{-25}$~J. In the region of atmospheric transparency a radio
frequency $\gamma$ can easily escape to the dark sky.

We stress that we do not investigate a new theory which replaces the
old one in a tricky way. The formalism just uses and extends the well
established quantum dynamics. Quantum dynamics allows to calculate
amplitudes between an initial and final state with convincing precision
not available in any macroscopic field of physics. It is taken as
the underlying theory. No arbitrary scale limiting its validity~\cite{ghirardi1986unified}
is introduced. That such a two vector state concept can be considered
a self consistent, time symmetric interpretation of quantum theory
was also claimed by Aharonov and Cohen~\cite{aharonov2017twotime}
and others~\cite{morita2015einstein}. 

The central difficulty in this interpretation is to understand \textit{how
the causal classical physics can arise} in the symmetric two boundary
theory. To proceed we introduced~\cite{Bopp:2018kiw,Bopp:2016nxn}
two transition rules which prohibit manifest macroscopic backward
causation. They essentially state that there can be no post selection~\cite{aharonov2015can}
and that the a-causality discussed above can usually be ignored as
it requires phase correlations in macroscopically distant sources
and unusual not phase averaging observations.

In macroscopic physics there is a \textit{causal decision tree}. At
each branching a decision how the future evolves is made. The critical
point is to understand the option ,,not chosen``. In macroscopia
quantum phases (and some minute changes) are averaged out. With such
incomplete macroscopic knowledge on both boundaries separate, ,,chosen``
or ,,not chosen`` macroscopic paths can appear if the distant between
the boundaries is sufficiently huge. 

The apparent time direction of the decision tree originates in our
relative proximity to the initial time and our huge distance to the
final one. 

Even with the limited macroscopic knowledge about the present situation
a lot is known about the past (Knowledge about fundamental processes
like the formation of stars is powerful.). Our assumption is that
given the full initial and present state in all macroscopically reachable
details there is only one macroscopic evolution path reaching us. 

It is not easy for macroscopically different states to evolve to the
same macroscopic state. Of course there can be strange attractors~\cite{mccauley1994chaos}
following and hiding earlier expansion periods which could allow for
ambiguous macroscopic evolutions. The assumption is that such exceptions
do not play any significant role in our mostly rather empty universe. 

The situation is different with the really distant final state. Here
a complete knowledge of the macroscopic boundaries still allows for
multiple coexisting paths. Our postulate is now that if the exact
quantum boundary conditions with all their given phases e.c.t. are
implemented these ambiguities vanish and the actually taken macroscopic
path is determined. In a classical consideration this selection is
mistaken to happen at the instance of the branching and it appears
that such decisions affects the choice of the future path.

This picture of the universe is perfectly consistent but there is
one annoying point. Self organization with an intrinsic time arrow
seems to play a central role also in the evolution of the universe.
This time arrow is not available in the final state. The concept is
that it suffices if these decisions encoded in the final state are
effectively random just like the decisions in the usual theory with
jumps. But still, to have the final state as a somehow external entity
which at least in principle decides everything touches basic scientific
principles. 

A reasonable scenario of the universe contains a big bang (at $t=0$),
a state of maximum extend (at $t=\frac{1}{2}T_{\mathrm{crunch}}$),
and a big crunch(at $t=T_{\mathrm{crunch}}$). As there is no intrinsic
time arrow the ,,forward moving`` world is formally symmetric to
the ,,backward moving`` one. Can the two state picture work in such
a universe~\cite{gell1994time,craig1996observation,davies1993time}? 

As above all macroscopic decision have to be encountered by a corresponding
loss and again an extreme miss match 
\[
\left\langle \mathrm{bang}|\mathrm{crunch}\right\rangle =10^{-huge}
\]
is needed. It does not mean fine tuning as there is a rich structure
which naturally rarely fits. It just requires that in the huge state
of maximum extent the probability of matching entanglements vanishes.
Essentially only one matching ,,border`` state should remain and
the $\left\langle \mathrm{bang}|\mathrm{(one\, effective\, final\, state)}\right\rangle $
picture should be resurrected.

A novel property is that the border state is now the result of evolutions.
It makes self organizing periods natural. 

What could be the relation between the forward and backward moving
part of the universe? As the border state is identical there have
to be some similarities. 

Both the boundary big bang and the boundary big crunch states could
be identical on a macroscopic level and the needed huge miss match
could just be due to details of their microscopic or  quantum dynamical
part. There are lots of phases available. 

Consider in such a scenario for the moment the border state only macroscopically.
Even with the macroscopically equal bang/crunch state many macroscopic
histories would contribute in both parts. 

The extremely extended border state is actually identical on a quantum
level for both sides. A somewhat daring assumption is now the this
fact could suffice to require equal macroscopic histories. This opens
an amusing option in which our macroscopia actually involves both
quantum epochs.

Let us consider this option in more detail. Consider the situation
with an electron wave with spin in the rightward direction the time
$t$ in the ,,forward moving`` world and an identical one at $T_{\mathrm{crunch}}-t$
in the ,,opposite moving`` world. A component $\propto\left\langle \mathrm{rightward}\,|\,\mathrm{upward}\right\rangle $
represents an upward intermediate state at $t+\epsilon$. We assume
this state to be uniquely traced in witnesses reaching the effective
final $T-\epsilon$ state. The component which reaches the same intermediate
state in the backward moving world at $T-t-\epsilon$ has an identical
size $\propto\left\langle \mathrm{rightward}\,|\,\mathrm{upward}\right\rangle {}^{CPT}$
. Averaging out unknown evolutions the probability of an upward spin
is therefore 
\[
P(\mathrm{sideward}\to\mathrm{upward})=|\left\langle \mathrm{sideward}\,|\,\mathrm{upward}\right\rangle |^{2}\,.
\]
The quadratic form of the Born rule is no longer a postulate but a
consequence of the concept. 

\noindent \begin{flushleft}
The seemingly statistical choice is no longer stored in a know-all-final
state but in an intrinsically matching state. The contribution to
the upward measurement is then: 
\begin{eqnarray*}
\left\langle bang\right| & U(T_{i},t)\cdot P_{up}\cdot\hspace{6.5cm}\\
 & U(t,T_{match})\cdot P_{match}\cdot U(T_{match},T_{crunch}-t)\\
 & \hspace{4.3cm}\cdot P_{up}\cdot U(T_{crunch}-t,T_{crunch}) & \left|crunch\right\rangle 
\end{eqnarray*}
 (with $P_{match}=1$ by continuity) and corresponding one for the
downward spin. 
\par\end{flushleft}

\noindent We consider now for both cases the central second line.
As argued the matching contributions are in both cases tiny say $10^{-huge}$
resp. $10^{-huge'}$. Given the extreme value of the exponents their
natural statistical variations $(\propto\sqrt{huge'})$ are large.
If $huge<huge'$ it therefore means $huge\ll huge'$ yielding a practically
exclusive dominance of its contribution. This ,,random`` decision
is no longer ,,würfelt`` (Einstein's term for dice) but it is consequence
of unknown ,,future`` evolutions.

The abolishment of the conventional time structure allows a curious
scenario in which we live with our wave function in the forward moving
world and with our conjugate function ,,eons apart`` in the tidily
correlated opposite moving one.

A tiny violation of CPT symmetry~\cite{vanTilburg:2016awx} could
be a signal for such a situation. 

Usually symmetry violations reflect properties of the Lagrangians
or, if preferred, simply of an asymmetric vacuum (Parity violation
needs a real grand unified theory like SO(10) broken down suitably
by the vacuum. For a natural vacuum mechanism for CP violation we
refer to~\cite{Bopp:2010um,Bopp:2011zz}). 

Such mechanism do not apply to CPT violation. CPT is a basic feature
of quantum dynamics or local field theories. Born's definition:
\[
Probability=Amplitude{}^{\mathrm{CPT}}\cdot Amplitude
\]
is manifestly invariant. 

In the bi-directional universe the big bang and big crunch state are
distinct on a quantum level and even their equality on a macroscopic
level might just hold in good approximation. Slightly different amplitudes
would then lead to a tiny CPT violation:

\[
Amplitude{}^{\mathrm{CPT}}\cdot Amplitude'\ne Amplitude\cdot Amplitude'^{\mathrm{\,\, CPT}}\,.
\]

Such a violation is a natural consequence of such asymmetric theories.
As a CPT asymmetry can also arise p.e. in elaborate non local field
theories~\cite{Greenberg:2002uu} it would strongly support but not
prove the bi-directional universe.

We thank David Craig, Eliahu Cohen, José M. Isidro and Giacomo D’ariano
for helpful correspondence. \pagebreak{}

\bibliographystyle{plain}
\bibliography{literatur}

\begin{thebibliography}{10}

\bibitem{aharonov1964time}
Yakir Aharonov, Peter~G. Bergmann, and Joel~L. Lebowitz.
\newblock Time symmetry in the quantum process of measurement.
\newblock {\em Physical Review}, 134(6B):B1410, 1964.

\bibitem{aharonov2015can}
Yakir Aharonov, Eliahu Cohen, and Avshalom~C. Elitzur.
\newblock Can a future choice affect a past measurement's outcome?
\newblock {\em Annals of Physics}, 355:258--268, 2015.

\bibitem{aharonov2017twotime}
Yakir Aharonov, Eliahu Cohen, and Tomer Landsberger.
\newblock The two-time interpretation and macroscopic time-reversibility.
\newblock {\em Entropy}, 19(3), 2017.

\bibitem{2015arXiv151206689A}
Yakir {Aharonov}, Eliahu {Cohen}, and Tomer {Shushi}.
\newblock {Accommodating Retrocausality with Free Will}.
\newblock {\em ArXiv e-prints}, December 2015.

\bibitem{Argaman2010Bell}
Nathan Argaman.
\newblock Bell s theorem and the causal arrow of time.
\newblock {\em American Journal of Physics}, 78(10):1007--1013, 2010.

\bibitem{balzer2002relaxationless}
Christoph Balzer, Thilo Hannemann, Dirk Rei{\ss}, Christof Wunderlich, Werner
  Neuhauser, and Peter~E. Toschek.
\newblock A relaxationless demonstration of the quantum zeno paradox on an
  individual atom.
\newblock {\em Optics communications}, 211(1):235--241, 2002.

\bibitem{block1991quantum}
Ellen Block and Paul~R. Berman.
\newblock Quantum zeno effect and quantum zeno paradox in atomic physics.
\newblock {\em Physical Review A}, 44(3):1466, 1991.

\bibitem{bohr1935can}
Niels Bohr.
\newblock Can quantum-mechanical description of physical reality be considered
  complete?
\newblock {\em Physical review}, 48(8):696, 1935.

\bibitem{bopp2013werner}
Fritz Bopp.
\newblock {\em Werner Heisenberg und die Physik unserer Zeit}.
\newblock Vieweg-Verlag, reprinted 2013, 1961.

\bibitem{Bopp:2010um}
Fritz~W. Bopp.
\newblock {Novel ideas about emergent vacua}.
\newblock {\em Acta Phys. Polon.}, B42:1917, 2011.

\bibitem{Bopp:2011zz}
Fritz~W. Bopp.
\newblock {Novel ideas about emergent vacua and Higgs-like particles}.
\newblock {\em Nucl. Phys. Proc. Suppl.}, 219-220:259--262, 2011.

\bibitem{Bopp:2016nxn}
Fritz~W. Bopp.
\newblock {Time Symmetric Quantum Mechanics and Causal Classical Physics}.
\newblock {\em Found. Phys.}, 47(4):490--504, 2017.

\bibitem{Bopp:2018kiw}
Fritz~W. Bopp.
\newblock {Causal Classical Physics in Time Symmetric Quantum Mechanics}.
\newblock In {\em {Proceedings of the 4th International Electronic Conference
  on Entropy and Its Applications, Basel, Switzerland, 2017. DOI:}}, 2018.

\bibitem{brown1957interferometry}
Robert~Hanbury Brown and Richard~Q. Twiss.
\newblock Interferometry of the intensity fluctuations in light. i. basic
  theory: the correlation between photons in coherent beams of radiation.
\newblock In {\em Proceedings of the Royal Society of London A: Mathematical,
  Physical and Engineering Sciences}, volume 242, pages 300--324. The Royal
  Society, 1957.

\bibitem{craig1996observation}
David~A. Craig.
\newblock Observation of the final boundary condition: Extragalactic background
  radiation and the time symmetry of the universe.
\newblock {\em annals of physics}, 251(2):384--425, 1996.

\bibitem{cramer1986transactional}
John~G. Cramer.
\newblock The transactional interpretation of quantum mechanics.
\newblock {\em Reviews of Modern Physics}, 58(3):647, 1986.

\bibitem{davies1993time}
Paul~C.W. Davies and Jason Twamley.
\newblock Time-symmetric cosmology and the opacity of the future light cone.
\newblock {\em Classical and Quantum Gravity}, 10(5):931, 1993.

\bibitem{de1953mechanique}
Olivier~Costa de~Beauregard.
\newblock M{\'e}chanique quantique.
\newblock 1953.

\bibitem{einstein1935can}
Albert Einstein, Boris Podolsky, and Nathan Rosen.
\newblock Can quantum-mechanical description of physical reality be considered
  complete?
\newblock {\em Physical review}, 47(10):777, 1935.

\bibitem{everett1957relative}
Hugh Everett~III.
\newblock "{R}elative state" formulation of quantum mechanics.
\newblock {\em Reviews of modern physics}, 29(3):454, 1957.

\bibitem{Gambini:2015zda}
Rodolfo Gambini and Jorge Pullin.
\newblock {The Montevideo Interpretation of Quantum Mechanics: a short review}.
\newblock 2015.

\bibitem{gell1994time}
Murray Gell-Mann and James~B. Hartle.
\newblock Time symmetry and asymmetry in quantum mechanics and quantum
  cosmology.
\newblock {\em Physical origins of time asymmetry}, 1:311--345, 1994.

\bibitem{ghirardi1986unified}
Gian~Carlo Ghirardi, Alberto Rimini, and Tullio Weber.
\newblock Unified dynamics for microscopic and macroscopic systems.
\newblock {\em Physical Review D}, 34(2):470, 1986.

\bibitem{Greenberg:2002uu}
Oscar~W. Greenberg.
\newblock {CPT violation implies violation of Lorentz invariance}.
\newblock {\em Phys. Rev. Lett.}, 89:231602, 2002.

\bibitem{griffiths1984consistent}
Robert~B. Griffiths.
\newblock Consistent histories and the interpretation of quantum mechanics.
\newblock {\em Journal of Statistical Physics}, 36(1-2):219--272, 1984.

\bibitem{joos2013decoherence}
Erich Joos, H~Dieter Zeh, Claus Kiefer, Domenico~JW Giulini, Joachim Kupsch,
  and Ion-Olimpiu Stamatescu.
\newblock {\em Decoherence and the appearance of a classical world in quantum
  theory}.
\newblock Springer Science \& Business Media, 2013.

\bibitem{kittel2005soft}
Wolfram Kittel and Eddi~A. De~Wolf.
\newblock {\em Soft multihadron dynamics}.
\newblock World Scientific, 2005.

\bibitem{mccauley1994chaos}
Joseph~L. McCauley.
\newblock {\em Chaos, dynamics, and fractals: an algorithmic approach to
  deterministic chaos}, volume~2.
\newblock Cambridge University Press, 1994.

\bibitem{metzger2011bose}
Wes~J. Metzger, Tamas Nov{\'a}k, Tamas Cs{\"o}rg{\H{o}}, and Wolfram Kittel.
\newblock Bose-{E}instein correlations and the tau-model.
\newblock {\em arXiv preprint arXiv:1105.1660}, 2011.

\bibitem{morita2015einstein}
Kunihisa Morita.
\newblock Einstein dilemma and two-state vector formalism.
\newblock {\em Journal of Quantum Information Science}, 5(02):41, 2015.

\bibitem{PenroseRoad}
Roger Penrose.
\newblock {\em {The Road to Reality: A Complete Guide to the Laws of the
  Universe}}.
\newblock Alfred A. Knopf, Inc, 2005.

\bibitem{plenio2008dephasing}
Martin~B. Plenio and Susana~F. Huelga.
\newblock Dephasing-assisted transport: quantum networks and biomolecules.
\newblock {\em New Journal of Physics}, 10(11):113019, 2008.

\bibitem{price2012does}
Huw Price.
\newblock Does time-symmetry imply retrocausality? {H}ow the quantum world says
  maybe?
\newblock {\em Studies in History and Philosophy of Science Part B: Studies in
  History and Philosophy of Modern Physics}, 43(2):75--83, 2012.

\bibitem{Vaidman:2014noa}
Lev Vaidman.
\newblock {Quantum Theory and Determinism}.
\newblock {\em Quant. Stud. Math. Found.}, 1(1-2):5--38, 2014.

\bibitem{vanTilburg:2016awx}
Jeroen van Tilburg.
\newblock {Measurements of CPT Violation at LHCb}.
\newblock In {\em {Proceedings, 7th Meeting on CPT and Lorentz Symmetry (CPT
  16): Bloomington, Indiana, USA, June 20-24, 2016}}, pages 73--76, 2017.

\bibitem{wheeler1949classical}
John~Archibald Wheeler and Richard~Phillips Feynman.
\newblock Classical electrodynamics in terms of direct interparticle action.
\newblock {\em Reviews of modern physics}, 21(3):425, 1949.

\bibitem{wigner1961remarks}
Eugene~P. Wigner.
\newblock Remarks on the mind body question, in "{The Scientist Speculates}".
\newblock {\em Heinmann, London}, 1961.

\bibitem{zeh2001physical}
Heinz~Dieter Zeh.
\newblock {\em The physical basis of the direction of time}.
\newblock Springer Science \& Business Media, 2001.

\end{thebibliography}

\end{document}